\documentclass[notoc]{JHEP3}

\title{A solution to the anisotropy problem in bouncing cosmologies}

\author{Valerio Bozza \email{valboz@sa.infn.it}\\
 Dipartimento di Fisica ``E. R. Caianiello",
Universit\`a di
Salerno, I-84081 Baronissi, Italy. \\
 Istituto Nazionale di Fisica Nucleare, Sezione di Napoli,
 Italy.}
\author{ Marco Bruni \email{marco.bruni@port.ac.uk}\\
 Institute of Cosmology and Gravitation, University
of Portsmouth, Dennis Sciama Building, Burnaby Road, Portsmouth
PO1 3FX, United Kingdom.}

\date{\today}

\abstract{ Bouncing cosmologies are often proposed as alternatives
to standard inflation for the explanation of the homogeneity and
flatness of the universe. In such scenarios, the present
cosmological expansion is preceded by a contraction phase.
However, during the contraction, in general the anisotropy of the
universe grows and eventually leads to a chaotic mixmaster
behavior. This would either be hard to reconcile with observations
or even lead to a singularity instead of the bounce. In order to
preserve a smooth and isotropic bounce, the source for the
contraction must have a super-stiff equation of state with
$P/\rho=w>1$. In this letter we propose a new mechanism to solve
the anisotropy problem for any low-energy value of $w$ by arguing
that high energy physics leads to a modification of the equation
of state, with the introduction of non-linear terms. In such a
scenario, the anisotropy is strongly suppressed during the high
energy phase, allowing for a graceful isotropic bounce, even when
the low-energy value of $w$ is smaller than unity. }

\begin{document}

\section{Introduction}

The classical problems of flatness and homogeneity that plague the
Big Bang cosmology can be successfully addressed by a phase of
accelerated expansion. In the standard inflationary scenario, the
expansion is driven by a scalar field rolling down its potential
\cite{inflation}. As a bonus, inflation predicts the existence of
quantum fluctuations in the initial vacuum state, leading to
primordial perturbations seeding the observed cosmic large-scale
structures \cite{inflpert}. These primordial fluctuations are
endowed with a nearly scale-invariant spectrum, in agreement with
observations of the cosmic microwave background \cite{CMB}.

On the other hand, the difficulties of embedding inflation within
a quantum gravity theory and the persistence of the initial
singularity in the inflationary scenario have motivated several
proposals of alternative cosmologies. There is a general consensus
on the existence of a high energy cut-off at the order of the
Planck scale, at which classical general relativity should be
replaced by a  quantum gravity theory. For example, in string
theory it has been suggested that the finiteness of the string
length should act as a natural cut-off for the curvature and the
energy density in the early universe \cite{BraVaf}. From this
point of view, the Big Bang singularity just represents the
outcome of the extrapolation of  general relativity beyond its
domain of applicability, whereas the  quantum gravity theory
should regularize this singularity, replacing it by a maximum in
the curvature and energy density of the universe. As a consequence
of the Big Bang regularization the existence of a contraction
phase before the Big Bang has been argued in several frameworks
\cite{PBB,LQG,Ekp,OtherBounce}. Following this hypothesis, the
universe should contract from initial conditions in a low energy
regime, evolving into a phase of higher and higher curvature,
until the high energy cut-off of the true quantum gravity theory
comes into play. This reverses the contraction into a standard
decelerated expansion, thus avoiding the general relativistic
singularity and replacing it by a cosmic bounce.

An appealing feature of bouncing cosmologies is that a contraction
phase would easily solve the flatness and horizon problems of the
classical Big Bang model without the need of inflation. However,
the contraction phase also inevitably leads to a problematic
growth of anisotropy. Einstein equations rapidly become anisotropy
dominated, while the contribution of matter becomes negligible,
leading to a ``velocity dominated" \cite{ELS,BS} singularity. This
typical outcome of general relativity \cite{Ashtekar} can only be
avoided if the energy density of the matter source grows faster
than the anisotropy. For a contracting universe, this happens if
the source behaves as super-stiff matter, i.e.\ if the ratio
$P/\rho=w$ of the pressure and the energy density is larger than
unity. For example, this condition can be realized by a scalar
field with a negative exponential potential, as in the Ekpyrotic
model \cite{Ekp}.

On the other hand, all sources with $w<1$ grow too slowly and are
finally taken over by anisotropies. When this happens, a mixmaster
scenario takes place, with the development of chaotic BKL
oscillations in the scale factors \cite{ELS,BS,Ashtekar,BKL,DL}.
It is not clear whether the quantum gravity phase would be able to
make its way out of the huge anisotropies generated during the BKL
oscillations \cite{DamHen}. The danger is that the universe
emerging after the bounce is too anisotropic to be acceptable.
Even worse, the universe could not manage to bounce, ending in an
anisotropic velocity dominated singularity.

The anisotropy problem in bouncing cosmologies represents a severe
shortcoming for all models with $w \leq 1$. This rules out
contractions driven by standard forms of matter, such as dust,
radiation or scalar fields with positive potentials. The search
for viable models of bouncing cosmologies is therefore restricted
to sources with $w>1$, which require scalar fields with peculiar
potentials or other more exotic forms of matter.

Is there a way to relax this constraint and recover models with
$w<1$ for the construction of viable contraction phases? As the
mixmaster behavior takes place at high energies, when strong
curvature effects dominate \cite{BS,Ashtekar,DL}, one possibility
is a modification of the effective equation of state (EoS) of
matter. Indeed, it may well be that the linear EoS  is  a valid
approximation only for sufficiently low densities and pressures
\cite{AnaBru}. Then, as the energy density increases, non-linear
terms in the EoS will eventually come into play. Thus, before
reaching any conclusion on the anisotropy problem in a contracting
phase, one should examine the effects of non-linearity of the EoS.

The purpose of this work is to show that the introduction of a
simple non-linear term in the EoS can alleviate or even completely
solve the anisotropy problem in bouncing cosmologies. A
suppression of the anisotropies by many orders of magnitudes can
be achieved, depending on the scale at which the non-linearity
starts to dominate. By a suitable implementation of this
mechanism, it becomes admissible to consider a low energy
contraction with $w<1$. As the simplest working example, we
consider here an EoS with a quadratic term.

\section{A toy model of fluid with Non-linear EoS}

Let us assume that the contraction era is dominated by a perfect
fluid with energy density  $\rho$ and pressure  $P$, and that
gravity in this phase is described by Einstein equations. The
general form of the EoS  including a quadratic correction is
\cite{AnaBru}
\begin{equation}
P=\alpha \rho + \epsilon \frac{\rho^2}{\rho_c}, \label{EOS}
\end{equation}
where $\alpha$ is a pure number that determines the low energy
EoS of the fluid, $\rho_c$ is the scale at which the
quadratic term becomes important and $\epsilon$ is the sign of the
quadratic correction\footnote{In general \cite{AnaBru}, one can of course  add to (\ref{EOS}) a constant $p_{0}$ term; however this is only relevant at low energies \cite{BBQ} and we don't need to consider it here.}. Here we will only focus on the case
$\epsilon=+1$.

One can think of Eq.\ (\ref{EOS}) as the second order truncation
of a series expansion in powers of $\rho$. In such case, the
linear term would represent the lowest order approximation to the
full EoS; $\rho_c$ can thus be interpreted as the non-linearity
scale of the microscopic theory of the fluid. It is also
interesting to note that that in several higher-order and quantum
gravity theories, and in particular in the brane scenario, the
corrections terms in the Einstein equations can be re-arranged as
quadratic corrections to the energy density (see e.g.\
\cite{Roy,brane} and refs.\ therein). However, in the brane
scenario the quadratic correction term appears in the effective
4-D Friedmann constraint equation, whereas  in the general relativity framework we adopt
here  the quadratic EoS affects the energy conservation  and the
Raychaudhuri equations (cf. \cite{AnaBru}). Hence, the dynamics of the
isotropisation mechanism in the two cases is not the same. In particular, in the brane scenario the effective quadratic term produces isotropy at high energy, but a singularity is unavoidable. Similar
corrections, but with the opposite sign, arise in Loop Quantum
Gravity, where the quadratic term is responsible for the bounce,
see e.g.\ \cite{LQC}.

In order to test the growth of the anisotropies, the simplest
approach is to consider a Bianchi I cosmology, which can be
described in terms of two dynamical quantities: the Hubble
expansion scalar $H$ and the traceless shear tensor
$\sigma_{\alpha\beta}$ (with $\alpha,\beta = 1, \ldots , 3$).
Introducing $\sigma^2 = \sigma_{\alpha\beta}
\sigma^{\alpha\beta}/2$, the energy conservation equation and
Einstein equations take the form (with $8\pi G/c^4=1$ )
\begin{eqnarray}
&& \dot \rho + 3H\left(\rho + P \right)=0\\
&& 3H^2-\sigma^2=\rho  \label{Friedman} \\
&& \dot H +H^2+ \frac{2}{3}\sigma^2 = -\frac{1}{6} \left(\rho+3P
\right) \\
&& \dot \sigma+3H \sigma =0.  \label{sigmadot}
\end{eqnarray}

Along with Eq.\ (\ref{EOS}), the above set of equations is closed
and can be solved in terms of the scale factor of the universe $a$
as
\begin{eqnarray}
&& \rho=\frac{(1+\alpha)\rho_c}{\left(\frac{a}{a_*}
\right)^{3(1+\alpha)} -1}  \label{rhosol}\\
&& \sigma^2=\sigma_i^2 \left(\frac{a}{a_i} \right)^{-6}\,,
\label{sigmasol}
\end{eqnarray}
where $a$ here is implicitly defined by $\dot{a}=H a$.

We are interested in a contraction phase driven by a source
satisfying the energy conditions, therefore we restrict to the
case $\alpha>-1$. The solution has two branches: here we focus on
the branch with $a>a_*$ with positive energy density.

For $a\rightarrow a_*$, the energy density (\ref{rhosol})
diverges, signaling the presence of a type III singularity (cf.\
\cite{AnaBru} and refs.\ therein). Of course we expect that
quantum gravity will dramatically change Eqs.\
(\ref{Friedman})-(\ref{sigmadot}) when $\rho$ reaches a cut-off
scale $\rho_M$, which can be expected to be of the order of the
Planck density $\rho_P= 4.6 \times 10^{113}$ J m$^{-3}$. At this
energy level, quantum gravity should intervene and drive the
universe across the bounce towards a standard decelerated
expansion phase. The details of this quantum phase are all to be
established, although several indications seem to point in the
right direction \cite{LQG,OtherBounce}. We will not deal with this
problem and just assume that the transition from contraction to
expansion occurs in this very high energy regime, when $\rho$  is
of the order of $\rho_M$.

The precise value of $a_*$ can be determined from the initial
conditions of the universe. In particular, if the universe starts
with a scale factor $a_i$ and energy density $\rho_i$, $a_*$ is
given by
\begin{equation}
a_*=a_i \left[(1+\alpha)\frac{\rho_c}{\rho_i}+1
\right]^{-1/3(1+\alpha)}.
\end{equation}

Similarly, once the cut-off scale $\rho_M$ is fixed, the scale
factor at the onset of the bounce is simply
\begin{equation}
a_M=a_* \left[(1+\alpha)\frac{\rho_c}{\rho_M}+1
\right]^{1/3(1+\alpha)}.
\end{equation}

\section{Anisotropy suppression}

The level of anisotropies in the universe can be estimated by
comparing the contribution of the shear term to that of the matter
source term in Eq.\ (\ref{Friedman}). It is clear that the
contraction of the universe will be driven by the larger term. So,
if we want to avoid an anisotropic approach to the bounce, we need
that the shear term be much smaller than the matter term at the
onset of the bounce. More explicitly, we wish to have
$\sigma^2_M/\rho_M \ll 1$.

In our simple model, it is very easy to extract the growth of the
anisotropies during the contraction phase. By simple algebra on
the exact solutions (\ref{rhosol})-(\ref{sigmasol}) and imposing
the hierarchy $\rho_i \ll \rho_c \ll \rho_M$, we can immediately
write down the following result
\begin{equation}
\frac{\sigma^2_M}{\rho_M} \simeq \frac{\sigma^2_i}{\rho_i}  \left(
\frac{\rho_c}{\rho_i} \right)^{\frac{1-\alpha}{1+\alpha}} \left(
\frac{\rho_c}{\rho_M} \right). \label{Result}
\end{equation}

The final (dimensionless) anisotropy fraction $\sigma^2_M/\rho_M$
is given by the initial anisotropy $\sigma^2_i/\rho_i$, multiplied
by an $\alpha$-dependent  growth factor arising in the low energy phase (when the
linear term in the EoS dominates) and a suppression factor ${\rho_c}/{\rho_M}$ arising
in the high energy phase (dominated by the quadratic term). Actually, the exponent of the growth factor depends on the barotropic
coefficient  $\alpha$ of the linear term in the EoS and is transformed into
an additional suppression factor when $\alpha>1$, as for
super-stiff matter, as it is well known. In this sense, Eq.\ (\ref{Result}) above recovers the well known suppression mechanism of super-stiff matter, and shows that this is an unnecessarily strong requirement provided that the EoS contains a quadratic term, giving rise to the ${\rho_c}/{\rho_M}$ suppression term.

The efficiency of growth in the linear phase and suppression in
the quadratic phase depend on the respective length of the two
phases. This is basically determined by the position of the
transition scale $\rho_c$. If we push it very close to the bounce
scale $\rho_M$, then the anisotropy suppression disappears and
only the growth due to the linear phase remains.  If instead we
push $\rho_c$ very close to $\rho_i$ assuming e.g.\ that the
entire pre-bounce is dominated by a purely quadratic EoS, then the
growth factor shrinks to one and the suppression factor becomes
huge.

Of course, the success of a quadratic EoS in retrieving an
isotropic universe crucially depends on the initial amount of
anisotropies in the initial conditions. If the universe is already
fairly isotropic, we do not need a very long quadratic phase in
order to wash out the shear generated during the low energy linear
phase. Then we can have $\rho_c$ fairly close to $\rho_M$. On the
other hand, if the universe starts in a very anisotropic state, we
need a longer quadratic phase, in order to ensure that the
universe does not enter a mixmaster regime.

It might now be useful to make a numerical example with some
familiar quantities, just to illustrate the effectiveness of the
quadratic EoS in the isotropisation of the universe in a given
case. First, we have to choose an initial value $\rho_i$ for the
energy density at the beginning of the contraction phase. There is
no universally accepted value for this quantity, since any
bouncing cosmology model makes a different hypothesis on the
initial state. Whatever the initial state of the universe, it must
be built up so as to satisfy some minimum consistency
requirements, in order to generate a viable cosmology. These
requirements can help us picking a representative value for the
initial energy density. For example, let us consider the solution
of the flatness problem. If one supposes that curvature and energy
density are comparable at the beginning of the contraction phase,
the flatness problem is solved only if the contraction starts with
$\rho_i<\rho_0$, where $\rho_0$ is the present energy density. By
choosing the present value of the radiation density,
$\rho_i=\rho_{\gamma,0}=4.2 \times 10^{-14}$ J m$^{-3}$, the
flatness problem is thus marginally solved. We will use this value
for numerical estimates and discuss the dependence of the final
results on this choice. As for the quantum gravity cut-off, we
simply assume it to be exactly at the Planck scale
$\rho_M=\rho_P$. Finally, the most important quantity is the
transition scale $\rho_c$. We might want it to lie at energies at
least higher than the nucleosynthesis, or even higher than the
baryogenesis, in order not to spoil standard cosmological results.
Then, we fix $\rho_c$ at the energy density of a radiative fluid
with the temperature $T_c=1$ TeV, which corresponds to
\begin{equation}
\rho_c=1.4 \times 10^{49} \left( \frac{T_c}{1 ~\mathrm{TeV}}
\right)^{4} \mathrm{J} ~ \mathrm{m}^{-3}.
\end{equation}

Assuming that the fluid dominating the contraction is pure
radiation, with $\alpha=1/3$, the anisotropy growth from the
initial state to the bounce is
\begin{equation}
\frac{\sigma^2_M/\rho_M}{\sigma^2_i/\rho_i} \simeq 5.4 \times
10^{-34} \left( \frac{T_c}{1 ~\mathrm{TeV}} \right)^{6} \left(
\frac{\rho_{\gamma,0}}{\rho_i} \right)^{1/2}. \label{ResRad}
\end{equation}

In this setup the initial shear is suppressed by 34 orders of
magnitude with respect to the dominant component energy density in
the course of the pre-bounce phase. For comparison, if we assume
no transition to a quadratic regime, the anisotropy grows by 63
orders of magnitude in a radiation-dominated contraction from
$\rho_{\gamma,0}$ to $\rho_P$.

With the values just used for this example, we can easily
calculate that the suppression would perfectly balance the growth
if we chose $T_c=2.5 \times 10^5$ TeV. The same would happen if we
kept $T_c=1$ TeV and set the initial energy density at $\rho_i= 3
\times 10^{-67} \rho_{\gamma,0}$.

Eq.\ (\ref{ResRad}) is for a radiation-dominated contraction,
which is the easiest case to realize, because all particles tend
to become relativistic as the temperature increases. However, a
dust-dominated contraction would be particularly interesting from
the point of view of cosmological perturbations. In fact, it is
well-known that the constant mode of the curvature perturbation
would develop a scale-invariant spectrum \cite{DualWands} and that
this would be easily transferred to the constant mode of the
post-bounce without invoking any exotic mechanism at the bounce
\cite{FinBra}. Unfortunately, a contraction dominated by
non-relativistic particles would sooner or later become
relativistic and thus replaced by a radiation-dominated one.
Non-relativistic particles can be mimicked by a scalar field with
a positive exponential potential, but the tracking solution is a
repeller in a contracting universe \cite{HeaWan}.
 It is interesting to note that Eq.\ (\ref{EOS}) with $\alpha=0$ can be
obtained by a K-essence model with the Lagrangian
\begin{equation}
\mathcal{L}=\rho_c\left[C \left(\frac{\chi}{\rho_c} \right)^{1/4}
\pm 1 \right]^2,
\end{equation}
where $C$ is a dimensionless parameter and $\chi=(\dot \phi)^2/2$.

Now, assuming that a stable dust-like contraction exists, we want
to test the efficiency of the quadratic correction in solving the
anisotropy problem with $\alpha=0$. Keeping $\rho_M=\rho_P$,
$\rho_i=\rho_{\gamma,0}$, $T_c=1$ TeV, and setting $\alpha=0$, Eq.
(\ref{Result}) becomes
\begin{equation}
\frac{\sigma^2_M/\rho_M}{\sigma^2_i/\rho_i} \simeq 10^{-2} \left(
\frac{T_c}{1 ~\mathrm{TeV}} \right)^{8} \left(
\frac{\rho_{\gamma,0}}{\rho_i} \right). \label{ResDust}
\end{equation}

Still we have a small suppression of the anisotropies, which is a
considerable step beyond the huge growth of anisotropies in a pure
dust contraction with $P\simeq 0$. This result can be improved by
lowering $T_c$. Otherwise, one has to assume that the initial
level of anisotropies is sufficiently low from the beginning.
Keeping anisotropies under control is essential if one wants to
generate a viable primordial spectrum of perturbations by a
dust-like contraction. It is easy to show that once the
perturbation mode is outside the horizon the spectrum of the
curvature perturbation is not affected by the presence of
non-linearities in the equation of state. Therefore, the
anisotropies must be largely negligible at least until the horizon
exit of the modes of interest for CMB observations and the
formation of large-scale structures. Since the horizon exit of
these modes occurs during the low energy linear phase,
anisotropies in the initial state must be very low in any case.

\section{Conclusions}

In this work we have clearly demonstrated that the inclusion of
non-linear terms in the EoS of the matter source driving the
contraction of a bouncing cosmology model has a huge benefic
effect on the suppression of the anisotropies. We have explicitly
considered a quadratic correction, but it is easy to imagine (and
we have explicitly checked) that more general non-linear terms
would yield a completely analogous result.

It can be objected that if the quadratic term in the EoS is
included in the source driving the post-bounce expansion as well,
then it would act in the opposite way, by temporarily enhancing
anisotropies. However, in general, one does not expect the nature
of the dominant source to be the same before and after the bounce.
Then, if there is any post-bounce non-linear regime, it is
sufficient that the transition to the standard linear regime
occurs at a higher temperature than the pre-bounce transition. In
such a way, the post-bounce would be protected from the return of
anisotropies.

Using this mechanism, it is possible to recover sources with EoS
coefficient $w<1$ as possible candidates for driving a pre-bounce
contraction. In fact, the approach to the bounce is no longer
menaced by the establishment of chaotic BKL oscillations. Models
of bouncing cosmologies can then safely use e.g.\ standard
radiative sources for the pre-bounce contraction without worrying
about anisotropies. Furthermore, this mechanism is particularly
interesting for any tentative revival of dust-like sources as
possible candidates for driving pre-bounce contractions. By
introducing non-linear corrections, one of the most severe
shortcomings preventing the development of bouncing cosmological
models can be finally overcome.

\begin{acknowledgments}
VB acknowledges support for this work by MIUR through PRIN 2006
Protocol 2006023491\_003, by research funds of Agenzia Spaziale
Italiana, by funds of Regione Campania, L.R. n.5/2002, year 2005
(run by Gaetano Scarpetta), and by research funds of Salerno
University. MB research was supported by
the UK's Science \& Technology Facilities Council.
\end{acknowledgments}

\end{document}